\documentclass[fleqn,10pt]{wlscirep}
\usepackage[utf8]{inputenc}
\usepackage[T1]{fontenc}
\usepackage{lineno}

\usepackage{graphicx}%
\usepackage{multirow}%
\usepackage{amsmath,amssymb,amsfonts}%
\usepackage{amsthm}%
\usepackage{mathrsfs}%
\usepackage[title]{appendix}%
\usepackage{textcomp}%
\usepackage{manyfoot}%
\usepackage{booktabs}%
\usepackage{algorithm}%
\usepackage{algorithmicx}%
\usepackage{algpseudocode}%
\usepackage{listings}%
\usepackage{gensymb}
\usepackage{booktabs} 
\usepackage{pbox}
\usepackage{subfigure}
\usepackage{epsfig,endnotes}
\usepackage{epstopdf}
\usepackage{times}
\usepackage{color}
\usepackage{colortbl}

\usepackage{xspace}
\usepackage{enumitem}
\usepackage{multirow}
\usepackage{tikz}

\usepackage{soul} 
\usepackage{tabularx}

\usepackage{tcolorbox}

\usepackage{times}
\usepackage{latexsym}

\usepackage{enumitem}
\usepackage[normalem]{ulem}

\usepackage{microtype}

\usepackage{inconsolata}

\usepackage{fontawesome}
\usepackage{stfloats}
\usepackage{amsmath}

\definecolor{red}{RGB}{228, 26, 28}
\definecolor{green}{RGB}{77, 175, 74}

\usepackage[table,xcdraw]{xcolor}

\title{Semantic-Aware Advanced Persistent Threat
Detection Using Autoencoders on LLM-Encoded
System Logs}

\author[1]{Waleed Khan Mohammed}
\author[1]{Zahirul Arief Irfan Bin Shahrul Anuar}
\author[1]{Mousa Sufian Mousa Mitani}
\author[2,*]{Hezerul Abdul Karim}
\author[3,*]{Nouar AlDahoul}

\affil[1]{Faculty of Artificial Intelligence
and Engineering, Multimedia University
Cyberjaya, Malaysia}
\affil[2]{Centre for Image and Vision Computing, Centre of Excellence for Artificial Intelligence, Faculty of Artificial Intelligence and Engineering, Multimedia University, Cyberjaya, Selangor, Malaysia}
\affil[3]{Computer Science, New York University Abu Dhabi, UAE}

\affil[*]{Joint senior authors}

\affil[* ]{
Correspondence: hezerul@mmu.edu.my, nouar.aldahoul@nyu.edu}

\begin{abstract}
Advanced Persistent Threats (APTs) are among the most challenging cyberattacks to detect. They are carried out by highly skilled attackers who carefully study their targets and operate in a stealthy, long-term manner. Because APTs exhibit “low-and-slow” behavior, traditional statistical methods and shallow machine learning techniques often fail to detect them.

Previous research on APT detection has explored machine learning approaches and provenance graph analysis. However, provenance-based methods often fail to capture the semantic intent behind system activities.
This paper proposes a novel anomaly detection approach that leverages semantic embeddings generated by Large Language Models (LLMs). The method enhances APT detection by extracting meaningful semantic representations from unstructured system log data. First, raw system logs are transformed into high-dimensional semantic embeddings using a pre-trained transformer model. These embeddings are then analyzed using an Autoencoder (AE) to identify anomalous and potentially malicious patterns. The proposed method is evaluated using the DARPA Transparent Computing (TC) dataset, which contains realistic APT attack scenarios generated by red teams in live environments. Experimental results show that the AE trained on LLM-derived embeddings outperforms widely used unsupervised baseline methods, including Isolation Forest (IForest), One-Class Support Vector Machine (OC-SVM), and Principal Component Analysis (PCA).
Performance is measured using the Area Under the Receiver Operating Characteristic Curve (AUC-ROC), where the proposed approach consistently achieves superior results, even in complex threat scenarios. These findings highlight the importance of semantic understanding in detecting non-linear and stealthy attack behaviors that are often missed by conventional detection techniques.
\end{abstract}

\usepackage[numbers]{natbib}

\begin{document}

\flushbottom
\maketitle

\thispagestyle{empty}

\section{Introduction}

APTs, considered a major threat to information security, differ from common malware attacks in their stealth, rarity, and extended attacks conducted by sophisticated attackers~\cite{BERRADA2020401}. Attackers often resort to "living off the land" to persist on the network without getting detected, thereby using and further extending system tools such as PowerShell and Windows Management Instrumentation (WMI) to carry out their malicious tasks seamlessly, much like their normal administrative tasks. The result is that system logs have become a major source of forensics used to analyze and identify attacks. The raw amounts of system-generated logs continue to make it impossible to scan them to identify any discrepancies, thus causing the need to use automated tools to identify anomalies~\cite{mitre_lotl_powershell}.

Although log analysis is of utmost importance, there are some major challenges for traditional anomaly identification techniques. Traditional techniques like IForest~\cite{liu2008isolation}, One-OC-SVM~\cite{manevitz2001one} and PCA~\cite{abdi2010principal} depended mainly on the statistical properties or the token frequency. These "shallow" learning methods regard log entries as points in a data set or merely as a bag-of-words, without regard to the extensive semantic context contained in the messages issued by the system~\cite{chandola2009anomaly, du2017deeplog, 7883294}. For example, a statistical model may have difficulty distinguishing a legal password reset operation by a system administrator from a privileged escalation attack initiated by a hacker if the statistical probability of the words is relatively close in value.

In order to overcome these drawbacks, this research paper proposes a transition from the conventional statistical anomaly detection method to semantic anomaly detection. The approach of utilizing models such as \texttt{all-mpnet-base-v2}~\cite{song2020mpnet} aims not only to enhance generalization by creating high-dimensional semantic embeddings of raw, unstructured system logs, but also to improve pattern recognition, enabling subtler behavioral differences to be identified more effectively. Later, semantic embeddings are fed into deep reconstruction models, particularly autoencoders~\cite{hinton2006reducing}, to robustly capture normal system behavior. Reconstruction loss is used to track anomalies. Events that do not get a good reconstruction by the model are considered to imply some intrusion. This method can discover malicious activities not because they are infrequent, but because their semantic behavior is different from the normal one.

The paper contributions are as follows:
\begin{enumerate}
    \item 
    This work builds on previous research~\cite{11011912} by proposing an unsupervised anomaly detection framework that used a pre-trained transformer language models, such as \texttt{all-mpnet-base-v2}, instead of the earlier Bidirectional Encoder Representations from Transformers (BERT) variants~\cite{devlin2019bert}, combined with autoencoders to detect APTs through semantic differences.
    \item We perform a comparative evaluation against three known baselines such as IForest, OC-SVM, and PCA. The results show benefits of applying a pre-trained embedding model to complex log data.
    \item We utilized data from the DARPA TC project~\cite{BERRADA2020401,Transparent_Computing}. The experimental results show that the combination of  \texttt{all-mpnet-base-v2} and autoencoder achieves higher AUC-ROC scores than those of the traditional baselines.
\end{enumerate}

\section{Related Work}
Recent cybersecurity research advances have increasingly explored the integration of deep learning due to the limitations of traditional anomaly detection and signature-based defense mechanisms~\cite{aldahoul2021model,mutalib2024explainable,mutalib2025explainable}. Specifically, semantic understanding, cross-domain correlation, and contextual reasoning have become the main enablers for detecting stealthy and low-frequency APTs. One of the works leverages LLMs and unsupervised modeling to enhance APT detection, correlation, and investigation~\cite{11011912}.
They present the development of a semantic anomaly detection framework that includes a system provenance traces transformed into natural language and then embedded into dense vectors by transformer models like BERT, ALBERT, and MiniLM, followed by an autoencoder-based reconstruction for anomaly scoring. Their proposed method outperformed traditional baselines. However, their proposed framework underlines several challenges like computational overhead and real-time applicability.

In~\cite{11141466}, the authors introduce a new modular framework where different agents handle specific tasks, such as checking emails, analyzing logs, and scanning IP addresses. The framework also includes a result correlation module that combines outputs from these agents to detect multi-step attacks. By using LLaMA-3.3-70B~\cite{grattafiori2024llama} along with methods like retrieval-based generation and chain-of-thought prompting, the approach shows strong performance in producing meaningful, domain-level insights and effectively identifying potential security threats~\cite{11141466}.

In other works, authors propose a two-phased strategy integrating anomaly detection using graph structures with LLM-assisted analysis for constructing comprehensive APT attack scenarios from audit logs~\cite{aly2025ocr}. The proposed model uses One-Class Relational Graph Convolutional Networks for anomaly detection within subgraphs. The detection is based on both structural and behavioral characteristics. An LLM-assisted retrieval pipeline is then employed to construct error-free, human-readable attack stories that are consistent with the APT kill-chain model. In their experiments with the DARPA TC3, OpTC, and NODLINK corpora, they found that there is immense improvement over existing solutions for both accuracy and investigation quality. However, there is still a need for future focus on simulating concurrency attacks, generalization, and integration with traffic data.

\section{Methodology}

In this section, we describe the datasets used in this work. In addition, we discuss our proposed improvement over existing methods~\cite{11011912}, which we call the MPNet-AE APT detection framework.

\subsection{Datasets}
The experimental evaluation is performed on the basis of the DARPA TC datasets~\cite{Transparent_Computing}, which form a widely adopted set of benchmarks for APT detection research. We used 5DIR, CADETS, CLEARSCOPE, THEIA, and TRACE scenarios, each capturing fine-grained system-level activities across heterogeneous operating systems and attack campaigns~\cite{11011912}. These datasets contain fine-grained provenance information, including process executions, system calls, file accesses, and network communications, providing a realistic trace of enterprise-scale system behavior~\cite{Transparent_Computing}.

All datasets are ingested and normalized using the Automatic Detection of Advanced Persistent Threats (ADAPT)  project’s ingester, which provides a consistent schema representation across various scenarios~\cite{11011912}. To analyze the contribution of different behavioral dimensions, each dataset is decomposed into five distinct provenance contexts:
\begin{itemize}
    \item ProcessEvent (PE): events at the system and kernel level.
    \item ProcessExec (PX): it runs binaries and traces of commands and shows all the processes currently running at this point.
    \item ProcessParent (PP): it refers to parent-child process relationships.
    \item ProcessNetflow (PN): this module represents network communication behavior.
    \item ProcessAll (PA): a single and uniform view of all contexts.
\end{itemize}

This multiview decomposition allows the anomaly detection performance to be evaluated in fine-grained terms with respect to various system attributes. The table~\ref{tab:datasets_final_simplified} summarizes the dataset by illustrating its highly imbalanced nature with APT attacks.

One of the biggest challenges in the DARPA TC datasets is the fact that they are highly imbalanced, meaning that less than 0.004\% of the records are malicious APT actions among APT networks. This makes the datasets ideal for testing models that are semantic-aware and reconstruction-based.

\begin{table*}[!h]
\centering
\caption{Experimental datasets of DARPA's TC program used in our study. A dataset entry (columns 3 to 7) is described by the number of rows (processes) / the number of columns (attributes). For instance, with ProcessAll (PA) obtained from the second scenario of Linux, the dataset comprises 279,369 rows and 457 attributes, including 100 APT attacks (0.036\%).}
\label{tab:datasets_final_simplified}
\small
\renewcommand{\arraystretch}{1.3}
\setlength{\tabcolsep}{7pt} 
\begin{tabular}{|l||c|c|c|c|c|c|c|c|}
\hline
 & Size & \textit{$PE$} & \textit{$PX$} & \textit{$PP$} & \textit{$PN$} & \textit{$PA$} & \textit{$nb\_attacks$} & \%$\frac{nb\_attacks}{nb\_processes}$ \\ \hline

BSD & 265 MB & 224624 / 32 & 224146 / 136 & 223780 / 38 & 42888 / 63 & 224624 / 266 & 11 & 0.005 \\ \hline

Windows & 36.0 MB & 11151 / 31 & 11077 / 372 & 10992 / 81 & 329 / 126 & 11151 / 607 & 9 & 0.081 \\ \hline

Linux (1) & 3.63 GB & 282087 / 26 & 271088 / 141 & 263730 / 46 & 6580 / 6226 & 282103 / 6436 & 46 & 0.016 \\ \hline

Linux (2) & 465 MB & 270652 / 19 & 278363 / 160 & 279090 / 71 & 48048 / 210 & 279369 / 457 & 100 & 0.036 \\ \hline

Android & 18.1 MB & 12106 / 28 & 12106 / 45 & 24 / 12 & 1116 / 214 & 12106 / 296 & 13 & 0.107 \\ \hline
\end{tabular}
\end{table*}

\subsection{Method}

The MPNet-AE APT detection framework receives process and network activity data from the provenance database and converts them into semantic embeddings generated by a pre-trained language model \texttt{mpnet-base-v2}~\cite{mpnet-base-v2}, followed by an autoencoder trained for anomaly detection. 

\subsubsection{Data Provenance Graphs}
The proposed framework relies on a provenance database as the core component for recording system execution behavior at the process level~\cite{BERRADA2020401,Transparent_Computing}. A single record of this database corresponds to an activity of an individual process. It consists of metadata, such as the process identifier and the sequence of system events generated during its execution. Such events can include system calls, executed binaries, relationships among processes considering their creation, which creates parent-child links, and active network interactions~\cite{BERRADA2020401,Transparent_Computing}.

Taken together, these logs provide a summarize view of system and network behaviour over time. Rather than examining raw log entries in isolation, the provenance-based representation captures contextual relationships between activities that can be leveraged to carry out more meaningful behaviour analysis. To facilitate semantic learning, each provenance record is later transformed into a structured textual description suitable for processing by a language model.

\subsubsection{Textual Conversion and Embedding Generation}
To facilitate the semantic interpretation of the system behavior, every record of the provenance is generated in the format of a natural language description, capturing the observed activities of the process. This helps fill the gap between the system logs and a human-interpretable description of the behavior. The sentence generation technique adopted follows the approach generally used in semantic log representation and APT detection techniques.

Features will only be listed for each process if they are active. The system will depend on the contextual view, whether it is process events, binary execution, parent processes, network traffic, or all contexts combined, in order to construct meaningful sentences to describe process activities, their interaction with the system, and whether they communicated with the network.
This sentence construction, which takes into account the context, ensures that the sentences or text generated are more likely to be similar to those written by a human and are not just a list of keywords or representations based on frequency.

To overcome the limitations of token-based and manually designed log features, we used \texttt{all-mpnet-base-v2} model~\cite{song2020mpnet,mpnet-base-v2}, a sentence embedding model based on Microsoft MPNet. MPNet is a transformer pretrained with a combination of Masked Language Modeling (MLM) and Permuted Language Modeling (PLM). It inherits the advantages of BERT and XLNet and avoids their limitations. \texttt{all-mpnet-base-v2} yields richer contextual representations than MLM-only BERT models. Combined with sentence-level contrastive fine-tuning, this enables the generation of high-quality semantic sentence embeddings.

\texttt{all-mpnet-base-v2} model~\cite{song2020mpnet,mpnet-base-v2} was already fine-tuned on a dataset of 1B sentence pairs. The Training process used a contrastive learning objective by giving one sentence, the model learns to identify its true paired sentence from a set of randomly sampled alternatives~\cite{song2020mpnet,mpnet-base-v2}.
The textual descriptions created previously are coded in the form of 768 dimensional dense vectors using \texttt{all-mpnet-base-v2}. 
These vectors are further normalized to ensure stability during autoencoder training.

These semantic embeddings are the key input features to the anomaly detection model. Unlike manually engineered features, the embeddings contain high-level, behavioral semantic meanings, enabling the system to identify subtle anomalies that share similar surface-level patterns with normal behaviors.

\subsubsection{Autoencoder}
The autoencoder consists of two major pieces: the encoder and the decoder. The encoder takes the high-dimensional semantic embeddings and maps them to a lower-dimensional space, which learns the essential characteristics of normal behavior in this process. It then decodes this mapping to reproduce the semantic embedding~\cite{hinton2006reducing,11011912}. When the AE is trained, the error in reconstruction for normal instances is minimized. In turn, this implies that the AE performs extremely well in reconstructing benign instances and not so well in reconstructing malicious instances. This is the starting point for implementing the Anomaly detector.

To replicate the normal system behavior, a vanilla autoencoder~\cite{hinton2006reducing} is used in this work. This autoencoder works by replicating the embeddings, which helps it to understand the patterns from the normal system processes~\cite{11011912}. This process was done using the unlabeled malicious data.

\subsection{Experimental Setup}
The aim of the set of experiments carried out in this work is to investigate the efficiency of the MPNet-AE framework. The autoencoder was trained using the embeddings from the typical system processes not the malicious ones.
The trained autoencoder model was tested on a hybrid dataset that has both normal and anomalous data. The model’s performance was evaluated in terms of reconstruction error analysis and AUC-ROC scores to determine how well the model can separate normal activities from the attacks based on a certain a threshold as done in previous works~\cite{11011912}.

A reconstruction error, which was obtained after training an autoencoder, was employed to determine the anomaly score for each sample. A sample with a high reconstruction error will be more prone to anomalies, hence classified as an anomaly.
The classification boundary based on the value of the reconstruction error, determines whether the data points fall into the normal or anomalous class. This boundary was chosen based on the validation dataset.

The experiments involving MPNet embeddings and the autoencoder were run on an NVIDIA RTX 1080 Ti GPU with 16 GB of VRAM. Other baseline models, such as Isolation Forest, OC-SVM, and PCA, were run on the CPU.

\section{Results and Discussion}
In this section, we will present and discuss the results of experiments performed to detect anomalous attacks.

\label{sec:exp-eval}

\subsection{Semantic Embedding Strategy using MPNet}
\begin{figure}[!h]
    \centering
    \includegraphics[width=1\linewidth]{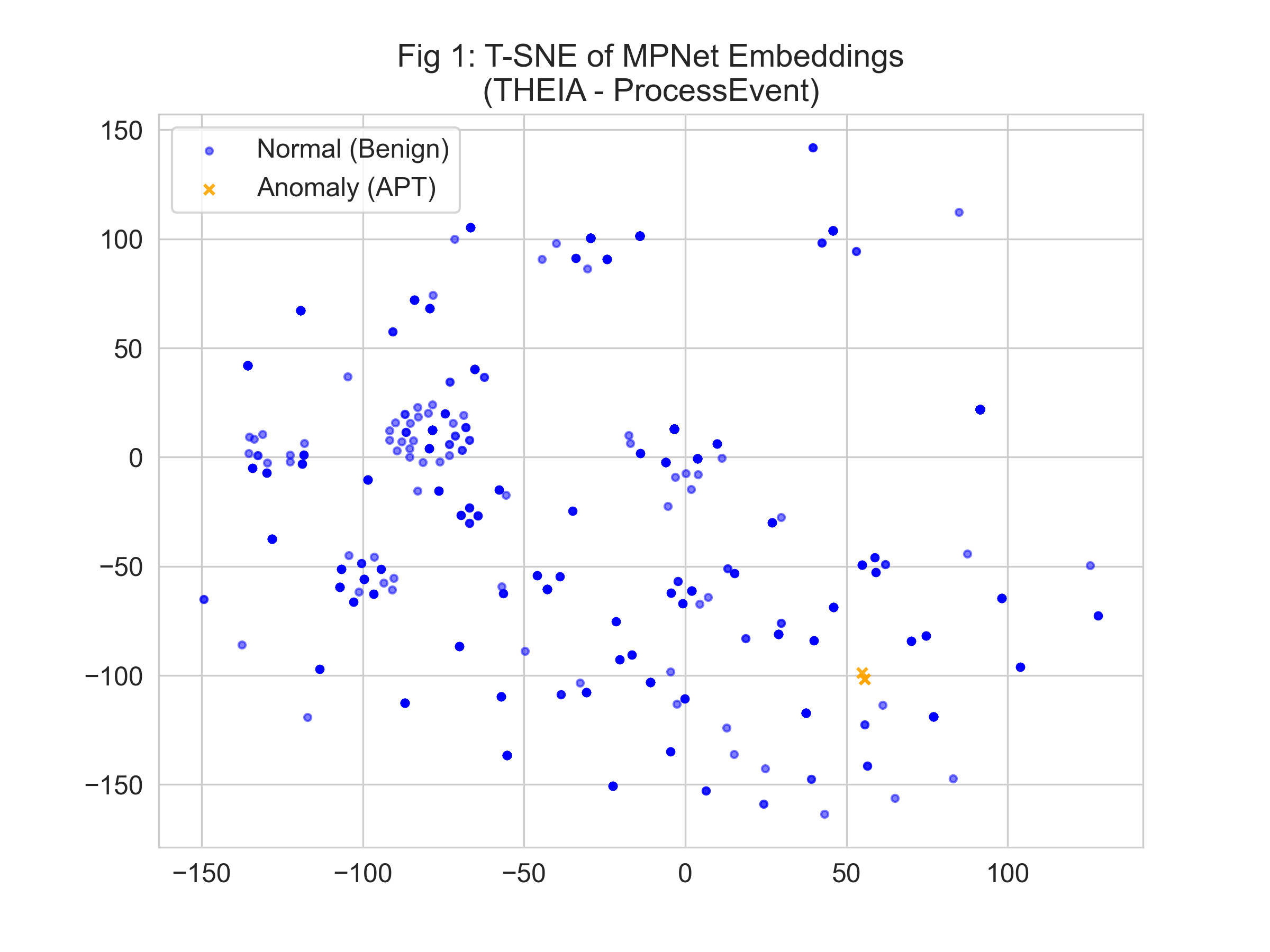}
    \caption{T-SNE visualizations of embeddings using \texttt{all-mpnet-base-v2}. Blue points (label 0) represent the normal data, whereas the orange crosses (label 1) represent the anomalies. In this example, data is represented for the PE dataset of the Linux OS.}
    \label{fig:tsne-theia-PE}
\end{figure}

We visualized the embedding of two attacks' classes using t-distributed Stochastic Neighbor Embedding (T-SNE)~\cite{maaten2008visualizing} for qualitative analysis.
As illustrated in Figure~\ref{fig:tsne-theia-PE}, normal process activities form compact clusters, while anomalous samples appear as isolated points, indicating effective semantic discrimination.  The observed patterns of clustering closely coincide with the high AUC-ROC of the autoencoder, which supports the design principles of the MPNet-AE framework.

The individual provenance records are encoded into a structured natural language representation containing process behavior in a readable form.  An example of a trace of a process execution is as follows:

\texttt{"Process 1054 started /bin/bash and connected socket 192.168.1.5:80 and changed /etc/passwd."}

These sequences are converted into tokens and processed by MPNet to produce 768-dimensional embeddings. Token representations are averaged using mean pooling to create a single fixed-size vector for each process instance. This method groups similar benign behaviors closely in the embedding space, while unseen or malicious APT activities are placed farther away, making anomalies easier to distinguish.

\subsection{Autoencoder Training and Detection Mechanism}
We fed the semantic representations into an autoencoder designed to act as a normal behavior model. The autoencoder uses a symmetric compression structure that gradually reduces dimensionality, allowing it to learn a compact representation of normal activity patterns.

\begin{figure}[!h]
    \centering
    \includegraphics[width=1\linewidth]{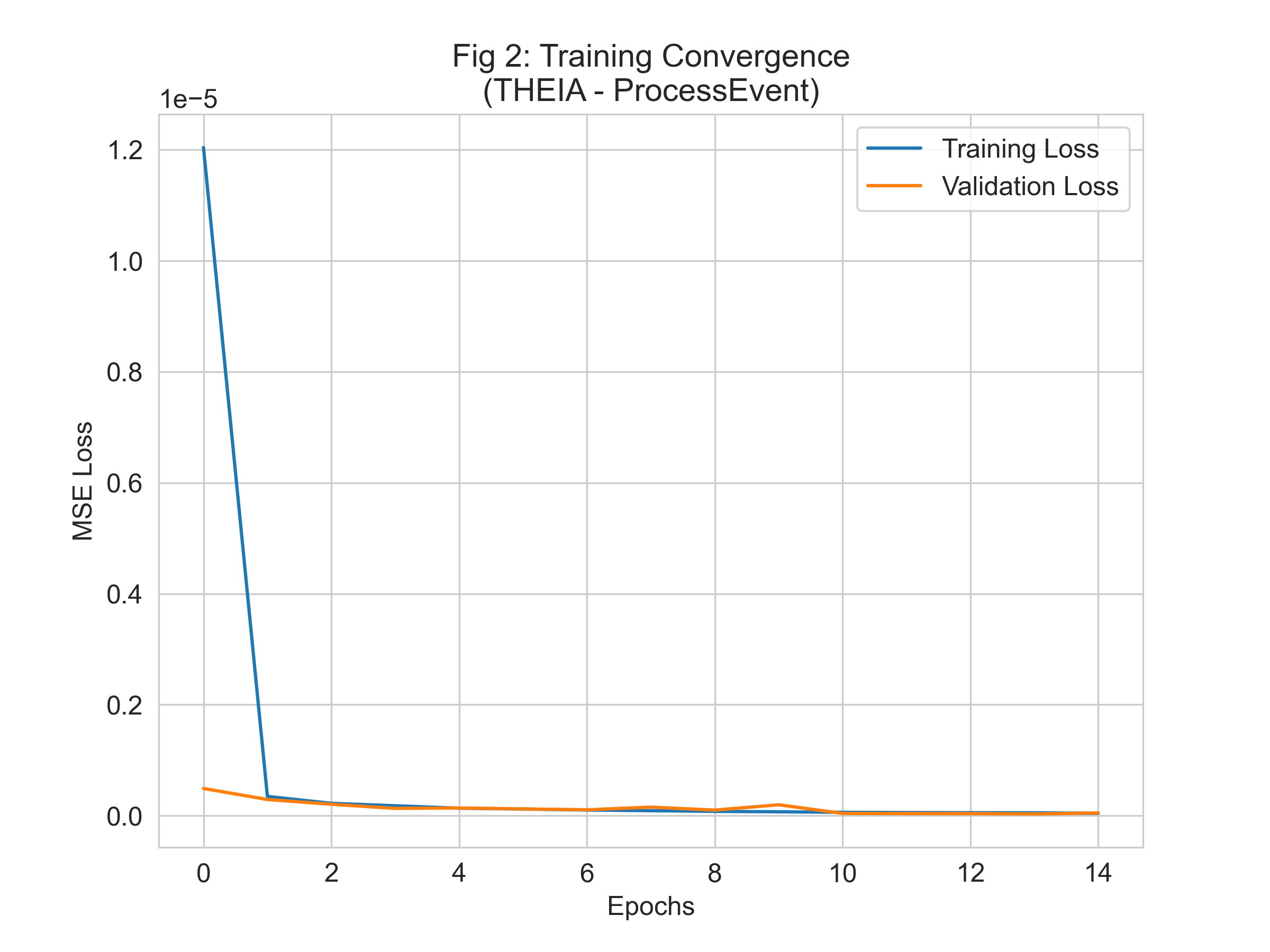}
    \caption{Training and validation loss curve for the autoencoder.}
    \label{fig:loss-curve}
\end{figure}

The model was fitted only on embeddings of normal behaviors using mean squared error (MSE) as the reconstruction loss metric. The number of epochs used in training was 15, training and validation loss curves had a rapid convergence at the beginning of the epochs and then reached a minimum. The similarity between the training and validation losses points to the strong generalization and to the insignificant overfitting. Figure~\ref{fig:loss-curve} shows the training and validation MSE curves over the epochs, indicating stable convergence and minimal overfitting.  

\begin{figure}
    \centering
    \includegraphics[width=1\linewidth]{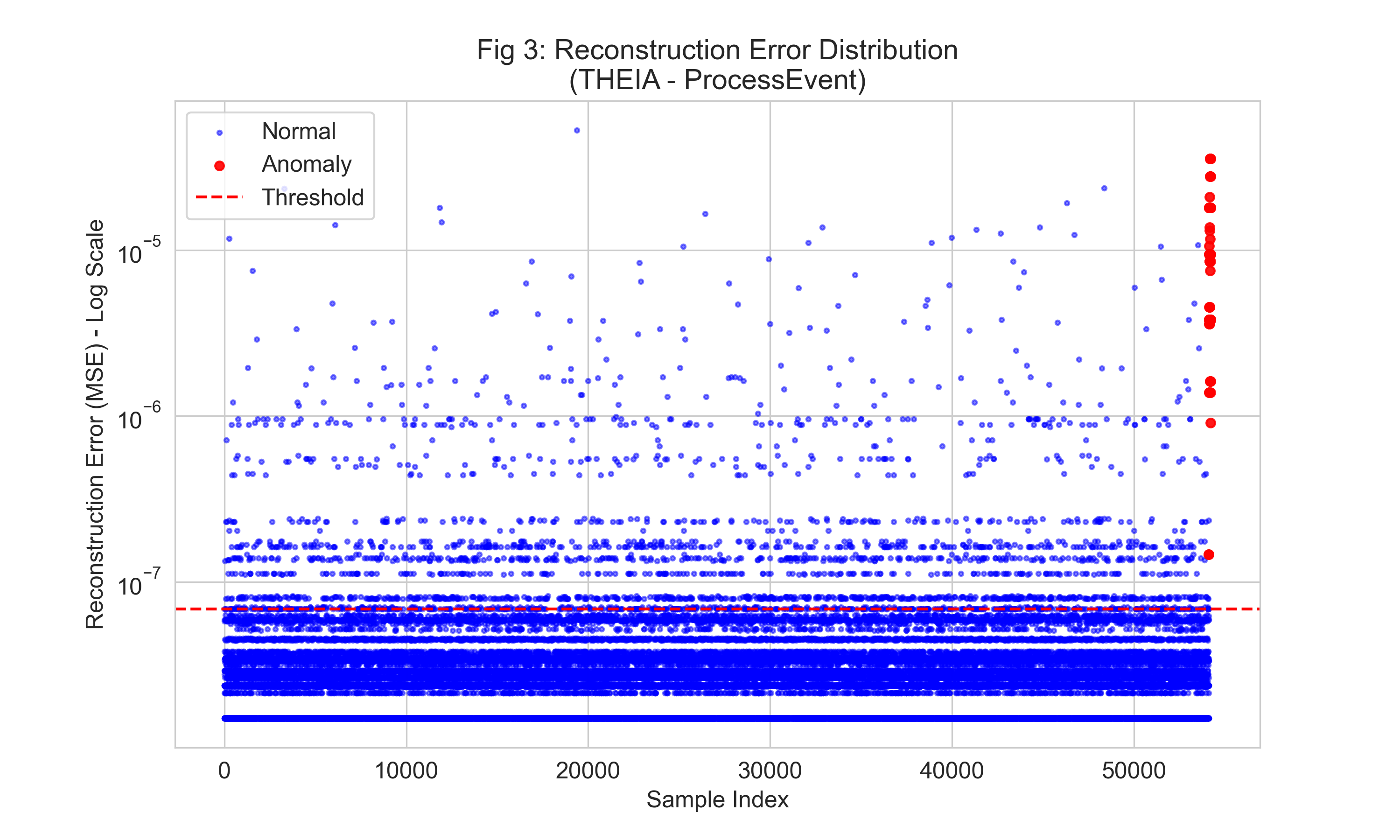}
    \caption{Scatter plot of the reconstruction errors in autoencoder for each sample. The red dashed line shows the anomaly detection threshold.}
    \label{fig:scatter-plot}
\end{figure}

When normal and abnormal samples combined in the testing time, the reconstruction error clearly separates them. Benign activities produced low reconstruction errors, while anomalous APT-related activities generated much higher errors because they differ from the learned patterns. This makes anomalies easy to detect using a threshold. As shown in Figure~\ref{fig:scatter-plot}, most normal samples have low reconstruction error, whereas anomalous activities exceed the threshold and are correctly identified.


\subsection{Performance Analysis}
\begin{figure}
    \centering
    \includegraphics[width=1\linewidth]{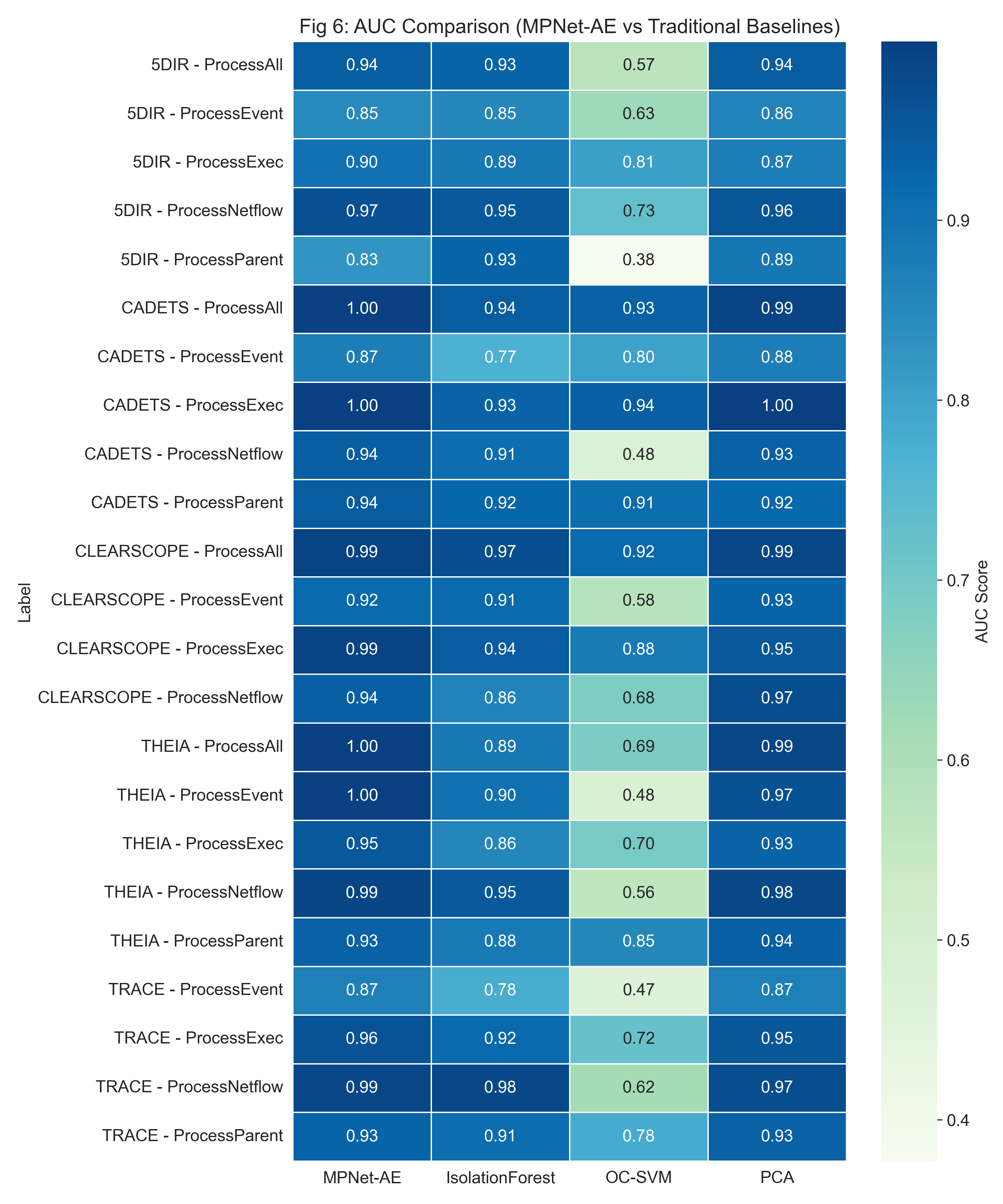}
    \caption{AUC heatmap comparing the performance of MPNet-AE, IForest, OC-SVM, and PCA across datasets from multiple OS (Windows, Linux, Android, and BSD).}
    \label{fig:heatmap}
\end{figure}

We compared the proposed MPNet-AE framework with three popular unsupervised anomaly detection baselines such as IForest, OC-SVM, and PCA using the AUC-ROC, which is quite effective in highly skewed data. As shown in Figure~\ref{fig:heatmap}, MPNet-AE consistently achieves higher AUC scores, demonstrating strong generalization across heterogeneous environments.

\begin{figure}[!h]
    \centering
    \includegraphics[width=1\linewidth]{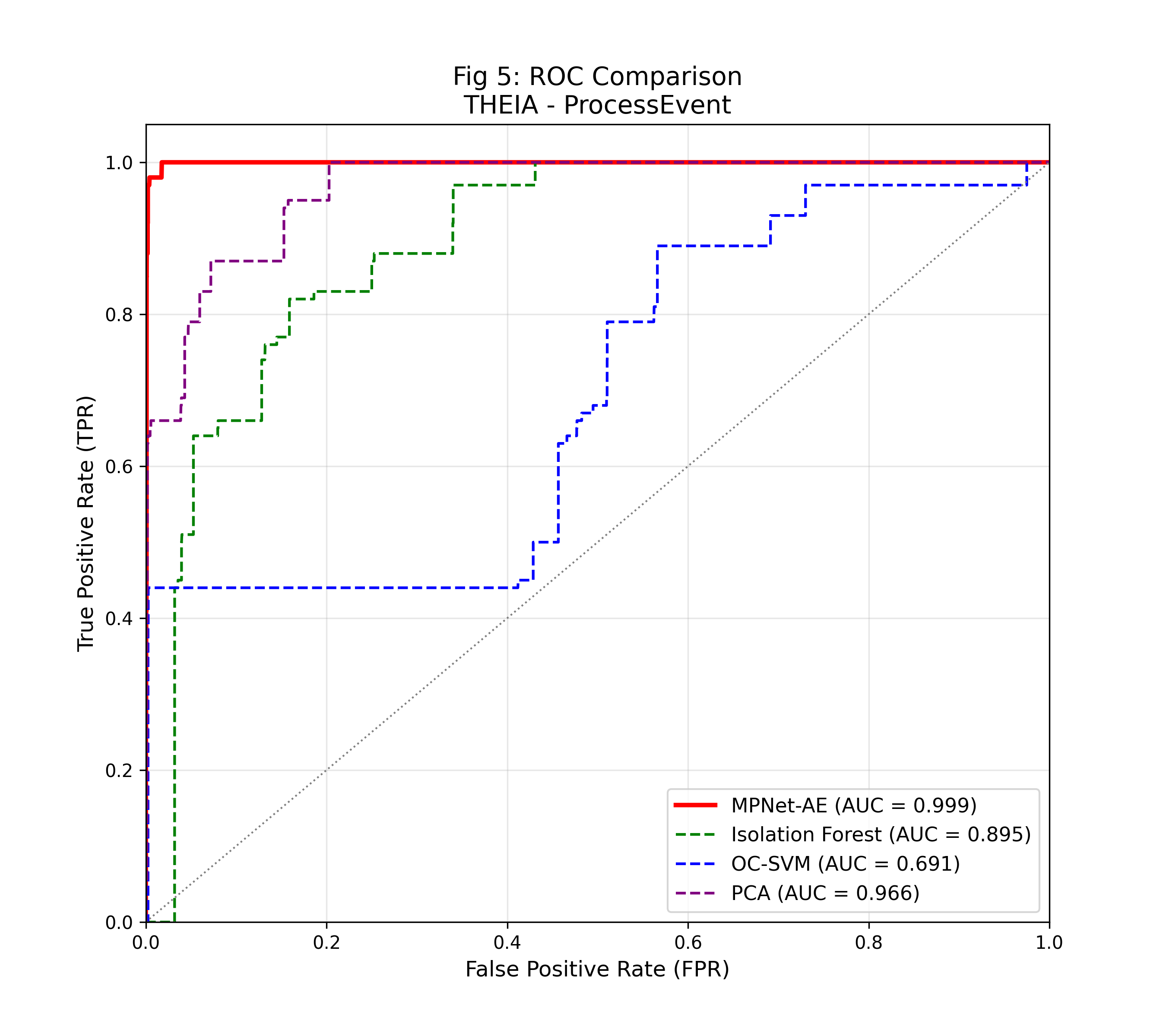}
    \caption{ROC curve for the MPNet-AE model.}
    \label{fig:roc}
\end{figure}

In all the datasets and provenance settings, MPNet-AE is always seen to be performing better than the traditional baselines. Specifically, it can reach AUC scores higher than 0.95 in 11 out of 23 experimental conditions as shown in Figure~\ref{fig:heatmap}. The ROC curve is shown in Figure~\ref{fig:roc}, where MPNet-AE achieves higher true positive rates across all thresholds, confirming its robustness under extreme class imbalance.

Since MPNet is a pre-trained model, no training was performed. On the contrary, the autoencoder requires end-to-end training because it has as a symmetric feed-forward neural network operating on 768-dimensional MPNet sentence embeddings. The encoder consists of two fully connected layers that progressively compress the input from 768 to 512 and then to a 128-dimensional latent representation, using ReLU activations to introduce non-linearity. The decoder mirrors this structure, reconstructing the input by expanding the latent vector through a 512-unit hidden layer back to the original 768-dimensional space. The model was trained using MSE reconstruction loss and optimized with the Adam optimizer at a learning rate of 0.001. Training was performed for 15 epochs with a batch size of 128.

For the OC-SVM model, the parameter nu was set to 0.01, indicating that approximately 99\% of the training data is assumed to be normal, while up to 1\% may be treated as anomalies. This choice makes the model conservative in labeling anomalies, focusing on learning a tight boundary around normal data. The radial basis function (RBF) kernel was used to capture non-linear patterns in the data, allowing the model to effectively separate normal behavior from outliers in complex feature spaces. Additionally, gamma was set to auto, meaning it is automatically determined based on the number of features, which helps balance model flexibility and generalization without manual tuning.

For the Isolation Forest model, the number of estimators was set to 100, meaning the algorithm builds 100 isolation trees. Each tree used binary splits, resulting in two branches per node, which is intrinsic to the isolation mechanism. The tree depth was not fixed manually. Instead, it was automatically determined by the max\_samples parameter which was set to auto, each tree was trained on a random subset of samples.

For PCA, we set n\_components = 0.95 means that PCA is configured to retain 95\% of the total variance in the original data rather than a fixed number of components. The algorithm automatically selected the minimum number of principal components required to explain at least 95\% of the variance, effectively reducing dimensionality while preserving most of the information.

The suggested solution provides a close to perfect detection rate in complicated cases like THEIA-ProcessEvent, which is significantly higher than isolated forest and the OC-SVM as shown in Figure~\ref{fig:heatmap}. Similarly, for TRACE, ProcessNetflow, with which patterns in the network traffic are very dynamic and noisy, MPNet-AE still has a high detection rate, but OC-SVM suffers from a drastic performance drop. This finding underlines the lack of effectiveness of traditional methods in handling high-dimensional semantic embeddings and nonlinear manifolds.

We found that PCA is competitive in fairly organized data like CADETS. PCA struggles most relative to MPNet-AE on complex relational views, especially ProcessExec and ProcessNetflow across datasets. PCA is rarely catastrophic, but it is systematically weaker than MPNet-AE in several settings.

Overall, MPNet-AE produces significant improvement compared to classical unsupervised baselines, proving that semantic representation learning with reconstruction-based modeling has a decisive advantage in the context of detection with APT in the process.

\section{Conclusion}

This paper introduces MPNet-AE, an unsupervised and semantically aware anomaly detection system designed for APT detection in provenance data. The approach captures contextual and semantic relationships among low-level system logs, information that traditional statistical features cannot represent, by converting logs into structured natural language and encoding them using the MPNet language model. The resulting embeddings are analyzed using reconstruction errors from an autoencoder trained solely on benign data, making the system effective at detecting stealthy and low-frequency attacks.

Tests on multiple DARPA TC datasets show that MPNet-AE consistently outperforms traditional unsupervised methods, achieving high AUC-ROC scores across different provenance scenarios, even when class distributions are highly imbalanced. Both quantitative results and visualizations confirm that malicious activities occupy distinct regions in the learned semantic space. This indicates that combining transformer-based representations with non-linear reconstruction learning is effective for detecting APTs.

Future research directions involve fine-tuning language models using logs specific to cybersecurity to further improve semantic discrimination. It is also worth exploring temporal and graph-based dependencies among processes to better detect multi-stage APT campaigns. Furthermore, we aim to examine other reconstruction models and adaptive online learning strategies to improve resilience in dynamic real-world settings. In addition, the expansion of the framework to live enterprise systems with the mechanisms of explainability will make it more practical in security operations.


\bibliography{sample}

\end{document}